# Leveraging Multiplexed Metasurfaces for Multi-Task Learning with All-Optical Diffractive Processors

Sahar Behroozinia[1] and Qing Gu[1,2*]

[1]Department of Electrical and Computer Engineering, North Carolina State University, Raleigh, 27695, USA
[2]Department of Physics, North Carolina State University, Raleigh, 27695, USA
*qgu3@ncsu.edu

## Abstract

Diffractive Neural Networks (DNNs) leverage the power of light to enhance computational performance in machine learning, offering a pathway to high-speed, low-energy, and large-scale neural information processing. However, most existing DNN architectures are optimized for single tasks and thus lack the flexibility required for the simultaneous execution of multiple tasks within a unified artificial intelligence platform. In this work, we utilize the polarization and wavelength degrees of freedom of light to achieve optical multi-task identification using the MNIST, FMNIST, and KMNIST datasets. Employing bilayer cascaded metasurfaces, we construct dual-channel DNNs capable of simultaneously classifying two tasks, using polarization and wavelength multiplexing schemes through a meta-atom library. Numerical evaluations demonstrate performance accuracies comparable to those of individually trained single-channel, single-task DNNs. Extending this approach to three-task parallel recognition reveals an expected performance decline yet maintains satisfactory classification accuracies of greater than 80% for all tasks. We further introduce a novel end-to-end joint optimization framework to redesign the three-task classifier, demonstrating substantial improvements over the meta-atom library design and offering the potential for future multi-channel DNN designs. Our study could pave the way for the development of ultrathin, high-speed, and high-throughput optical neural computing systems.

## 1. Introduction

Optical computing, especially through Optical Neural Networks (ONNs), has long been recognized for its potential to enhance computational speed and energy efficiency. The first optical implementation of neural networks in 1987 used optical components to emulate neuron configurations, sparking decades of research into optical neuromorphic technologies [1]. Recently, advancements in deep learning and photonic technology have revitalized interest in this field, enabling the development of scalable, ultra-fast, and energy-efficient ONNs [2-9]. Diffractive Neural Networks (DNNs) [10-22] are a type of ONN consisting of multiple spatially engineered transmissive diffractive layers. Utilizing light-matter interactions, these diffractive surfaces perform element-wise multiplication, with each 'pixel' acting as a 'neuron', interconnected through the physics of optical diffraction. The complex-valued transmission coefficient of each neuron serves as a trainable network parameter, systematically adjusted via an error back-propagation algorithm executed on a digital computer to perform a specific machine learning task.

On the other hand, metasurfaces – engineered two-dimensional arrays of subwavelength nanostructures – allow one to precisely manipulate optical properties such as phase, amplitude, and polarization, all through the adjustment of the size and shape of meta-atoms [23-27]. Over the past decade, this capability has revolutionized applications in several fields, including imaging and holography [28-31], sensing [32, 33], information processing [34-36], and quantum photonics [37, 38]. Their integration into DNNs has facilitated the development of advanced ultra-thin diffractive processors, which show promise for large-scale on-chip integration in future computing systems. Additionally, the capability of metasurfaces for multi-dimensional light modulation makes them ideal for constructing multi-channel, multi-functional computing devices. Several studies have demonstrated how multiplexing of various optical properties – such as wavelength, polarization, and angle of incidence [39-42] can be harnessed to develop compact, parallel optical computing systems capable of performing mathematical operations such as differentiation and integration within a single element.

Although most DNNs focus on performing a single machine learning task, the ability to handle multiple tasks within a single DNN is crucial for advancing toward more generalized artificial intelligence devices that are both high-speed and energy-efficient, for applications such as autonomous driving and machine vision. Recently, significant steps have been taken toward implementing versatile, multi-functional DNNs. For instance, one study introduced a reconfigurable metasurface-based pluggable DNN capable of switching between two tasks by altering its pluggable components [43]. Another approach experimentally demonstrated an on-chip two-task optical classifier using birefringent nanostructures and a polarization multiplexing scheme, despite limited performance with a single-layer architecture [44]. Additionally, a numerical investigation explored multi-wavelength parallel image recognition of more than two tasks, utilizing a joint optimization approach to adjust the height map of diffractive optical elements [45]. However, achieving high parallel classification accuracy in this case requires at least five diffractive layers and a large number of modulation elements per layer, which makes the device bulky. In short, while these developments mark important progress toward versatile, multifunctional diffractive processors, the full potential of multiplexed metasurfaces has yet to be fully explored in terms of utilizing physical parametric degrees of freedom to implement compact, highly parallel multi-task DNNs.

In this work, we rigorously investigate the potential of both polarization-multiplexed and wavelength-multiplexed metasurfaces in realizing DNNs capable of simultaneously classifying multiple inputs. Using our meta-atom library, we initially design a dual-channel Polarization-Multiplexed DNN (PM-DNN) and a dual-channel Wavelength-Multiplexed-DNN (WM-DNN), to simultaneously classify the MNIST and Fashion-MNIST (FMNIST) databases with high classification accuracies. Extending this approach, we introduce a tri-channel WM-DNN to perform three tasks, MNIST, FMNIST, and Kuzushiji-MNIST (KMNIST), in parallel. Numerical results demonstrate satisfactory outcomes, despite a moderate decline in accuracy for the two more challenging tasks of FMNIST and KMNIST, primarily due to increased task competition. To further enhance the system performance, we develop a novel end-to-end design methodology to redesign the tri-channel WM-DNN. This framework utilizes surrogate models to map the complex transmission responses of the meta-atoms to their structural parameters, which are then used in a joint training framework to optimize the network parameters for all three tasks simultaneously. This approach not only improves classification accuracy but also has the potential to train multi-channel DNNs capable of handling a large number of tasks in parallel, thereby enabling massively parallel, multifunctional neural architectures.

## 2. Design

Fig. 1(a) schematically shows the concept of metasurface-assisted Multiplexed DNNs (M-DNNs) for parallel optical classification. The system includes a multi-channel optical field with different targets encoded in specific light channels as the input layer, multiplexed metasurfaces as hidden layers, and a segmented detection plane for multi-channel detection functioning as the output layer. The two orthogonal polarization states x and y in the upper panel of Fig. 1(a) and the three wavelengths $\lambda_1$, $\lambda_2$, and $\lambda_3$ in the lower panel serve as independent channels (i.e., without any cross talk) for information processing. Adjusting the structural parameters of each meta-atom allows for spatially varying, channel-dependent transmission responses, enabling independent processing of the multi-input light. Each meta-atom in a designated polarization or wavelength state acts as an 'optical neuron', which is interconnected with the neurons in the subsequent layers through the physics of optical diffraction. According to the Rayleigh–Sommerfeld integral [46], the complex field of the $(l+1)$th layer of the M-DNN can be expressed as:

$$E_{p,\lambda}^{l+1}(x,y) = \iint_{-\infty}^{+\infty} E_{p,\lambda}^{l}(x',y') \cdot T_{p,\lambda}^{l}(x',y') \cdot \frac{z}{r'^2}\left(\frac{1}{2\pi r'} - \frac{1}{j\lambda}\right) exp\left(j\frac{2\pi r'}{\lambda}\right) dx' dy' \quad (1)$$

Where $E_{p,\lambda}^{l}(x',y')$ is the optical field irradiated to the $l$th layer, for the specific channel with polarization $p$ and wavelength $\lambda$. For the $l = 1$, $E_{p,\lambda}^{l}(x',y')$ represents the projected light of the target image. The term $z = z^{l+1} - z^{l}$ denotes the axial distance between successive layers, $r' = \sqrt{(x-x')^2 + (y-y')^2 + z^2}$ and $j = \sqrt{-1}$. $T_{p,\lambda}^{l}(x',y')$ is the spatially varying complex transmittance of the metasurface of the $l$th layer corresponding to the channel with polarization $p$ and wavelength $\lambda$, which is a function of the structural parameters of the meta-atoms. The multi-channel light is modulated by the metasurfaces, and the multi-dimensional transmission responses of the meta-atoms—i.e., their modulation coefficients—can be optimized to maximize the detected light intensity within the correct detection sub-region for each task, corresponding to its respective channel, thereby enabling optical multitask learning. It is important to note that while channel crosstalk does not occur as light diffracts through the air or substrate and is modulated by the metasurfaces, it does manifest at the detection plane due to the superposition of optical intensities. Employing polarization-selective and wavelength-selective filters in each sub-area can effectively mitigate crosstalk during intensity measurements across various channels, thereby enhancing the performance of the M-DNN. In this work, we assume the presence of such filters on the detectors.

To illustrate the capabilities of our M-DNN, we present several examples of multitask optical systems. As shown in Fig.1(b), these multiplexed metasurfaces consist of rectangular $TiO_2$ nanofins on a glass substrate with a fixed height $h$ and two independently tunable widths $w_x$ and $w_y$. When exposed to linearly polarized light, the asymmetric meta-units modulate the phase and amplitude of the incoming light in a polarization- and wavelength-dependent manner. By adjusting $w_x$ and $w_y$, the desired transmission responses for each channel are achieved, facilitating parallel multitasking across various polarization and wavelength incidences. The phase and amplitude of the complex transmission responses of the meta-atoms are modeled using COMSOL Multiphysics software, which utilizes the Finite Element Method (FEM). The nanofins are set with a fixed height of 600 nm and a period of 400 nm. Fig. 1(d-f) display the computed electromagnetic response of each unit cell under x- and y-polarizations for specific wavelengths of 450 nm, 550 nm, and 650 nm, with the nanofins' widths ranging from 60 nm to 350 nm in 5 nm increments.

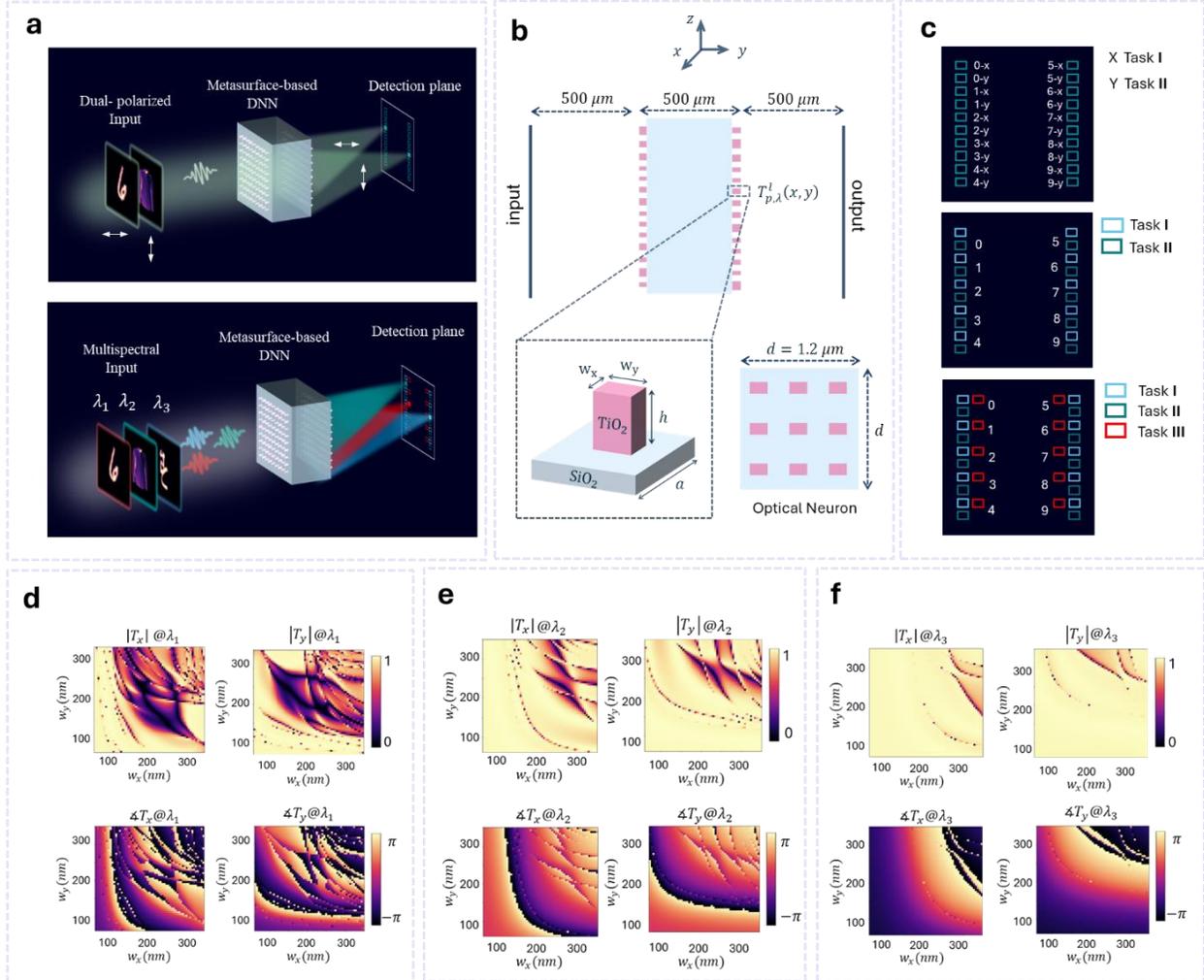

Fig. 1(a) Concept of metasurface-based M-DNN for performing multiple tasks. Top: Dual-channel PM-DNN. Bottom: Tri-channel WM-DNN. After the multi-input light, comprising various datasets and encoded in specific polarization (top panel) or wavelength (bottom panel) channels, passes through the doublet metasurface, it is focused onto the corresponding detection areas for each task's class, enabling parallel recognition. (b) Schematic of the M-DNN featuring a doublet metasurface. The axial distance between the layers is set at 500 μm. The metasurface consists of $TiO_2$ rectangular nanopillars on a glass substrate. The height of each meta-atom $h$ is fixed at 600 nm, while the widths, $w_x$ and $w_y$, vary from 60 nm to 350 nm; the periodicity of the unit cells, denoted as $a$, is 400 nm. Each meta-atom exhibits a polarization- and wavelength-dependent transmission response. A supercell, consisting of a 3×3 array of identical meta-atoms, serves as our optical neuron. Their channel-dependent transmission responses are optimized through the training process to perform multiple classification tasks. (c) Designated detection areas corresponding to each task category for the dual-channel PM-DNN (top), dual-channel WM-DNN (middle), and tri-channel WM-DNN (bottom). (d-f) Simulated values of the transmission amplitude ($|T_x|, |T_y|$) and phase ($\angle T_x, \angle T_y$) for different geometries, under x- and y-polarized light. The incident wavelengths for (d), (e), and (f) are 450 nm, 550 nm, and 650 nm, respectively.

All designs presented in this paper feature two hidden layers, i.e., multiplexed metasurfaces, each containing 210 × 210 optical neurons. Instead of utilizing a single unit cell, we intentionally adopted a supercell configuration consisting of a 3×3 array of identical meta-atoms, serving as our optical neuron, as depicted in Fig.1(b). This enlargement of the neuron size to 1.2 μm × 1.2 μm facilitates the feasible experimental realization of the designs using commercially available spatial light modulators and CCD cameras, whose smallest pixel sizes are in the range of a few microns. Therefore, utilizing a supercell structure instead of a single cell allows us to maintain the desired pixel size for efficient light encoding and detection, while also benefiting from the light modulation capabilities of meta-atoms at shorter periodicities, specifically 400 nm. Another critical parameter in the design of a DNN is the distance between successive layers. Since 500 μm is the most commonly available thickness for $SiO_2$ wafers, we have fixed the axial distance between the layers at 500 μm to support the ease of implementation in the future.

## 3. Designing M-DNNs using a meta-atom library

In this section, we verify the application of our metasurface-based M-DNNs in parallel multi-task classification utilizing our library of unit cell configurations—sourced from our simulations. We first design a dual-channel PM-DNN and a dual-channel WM-DNN to simultaneously classify two distinct datasets: the MNIST database (task I) and the Fashion-MNIST (FMNIST) database (task II), containing handwritten digits and fashion items, respectively. In the PM-DNN, MNIST and FMNIST data are encoded in x- and y-polarized light, respectively, both using a wavelength of 550 nm. In contrast, for the WM-DNN, the handwritten digits from task I are encoded at a wavelength of 450 nm, and the fashion products from task II at 550 nm, both under x-polarization. Fig. 1(c) illustrates the detection planes for the dual-channel PM-DNN and WM-DNN, where specific sub-areas are designated for each category, with the upper and lower regions corresponding to tasks I and II, respectively.

Training is performed by individually training four single-task DNNs, each specifically designed for a distinct polarization and wavelength corresponding to the channels of our PM-DNN and WM-DNN. Each DNN is tasked with performing a specific classification task. The transmission amplitudes of the neurons are set to unity, and the transmission phases $\varphi$ are the trainable parameters (phase-only DNNs). These phases are optimized via an error back-propagation algorithm to focus the output light intensity on the designated sub-area corresponding to each task's category. The top panels of Figs. 2(a) and 3(a) display the phase distributions of the hidden layers obtained from the training sessions for the dual-channel PM-DNN and WM-DNN, respectively, designed to classify MNIST and FMNIST datasets. Subsequently, for each location on the metasurfaces, we search within our existing unit cell library to select the optimal meta-atom structure that most accurately replicates the desired local phase shifts for both tasks (More details on the selection process can be found in the Materials and Methods section). The lower panels of Figs. 2(a) and 3(a) show the mismatch between the optimized phase shifts and those realized by the PM-DNN and WM-DNN, respectively, highlighting that the metasurfaces successfully reproduce the learned phase profiles with high fidelity. Numerical evaluations of the metasurface-based PM-DNN and WM-DNN using unseen data confirm the capabilities of both dual-channel classifiers, as they achieve classification results comparable to those of their individually trained single-task counterparts. The classification accuracies obtained are 97.72% for MNIST and

88.01% for FMNIST by the PM-DNN, and 97.19% for MNIST and 86.35% for FMNIST by the WM-DNN, respectively, as presented in Table 1.

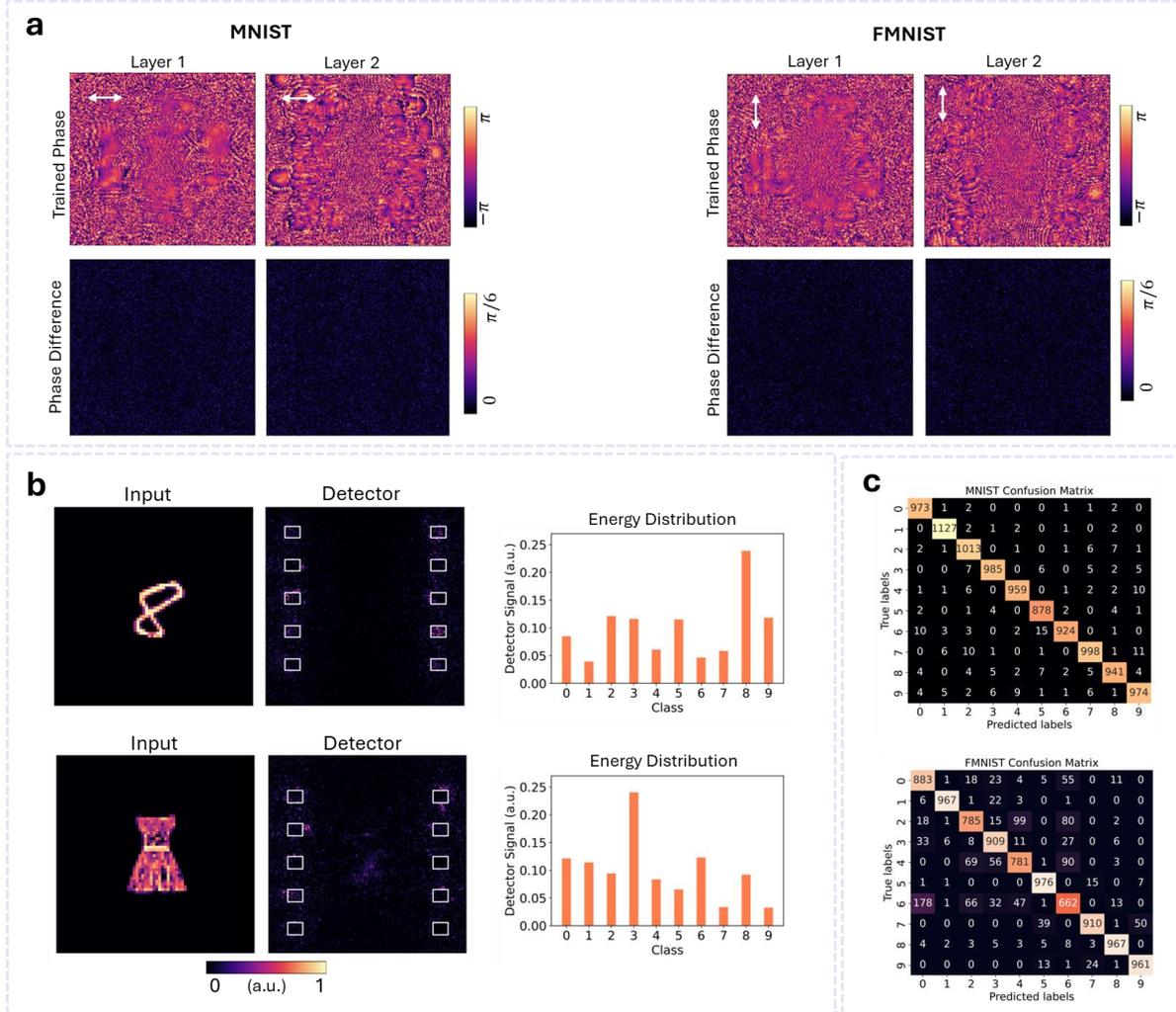

Fig. 2(a) (Top) Final phase distributions of the two hidden layers of the dual-channel PM-DNN under x- and y-polarized incident light with the incident wavelength set to 550 nm. (Bottom) The absolute phase difference between the desired phase and the phase realized by the meta-atoms at each pixel. (b) Exemplary results of simultaneously classifying a handwritten digit and a fashion item encoded in the x- and y-polarized light, respectively. Output intensity patterns and normalized energy distributions across the category sub-areas show the success of the PM-DNN in parallel two-task categorization. (c) Confusion matrices for MNIST and FMNIST processed by the PM-DNN, demonstrating its performance across individual classes, with average classification accuracies of 97.72% and 88.01% for Task I and Task II, respectively.

Figs. 2(b) and 3(b) demonstrate the performance of these M-DNNs in parallel recognition of a handwritten digit and a fashion item, where the digit '8' and the item 'dress' are processed by the PM-DNN, while '0' and 'trouser' are processed by the WM-DNN. For both two-task classifiers, the detector intensity patterns, and normalized output energy distributions clearly indicate that the correct sub-regions corresponding to Task I and Task II receive the maximum signal. The confusion matrices, which statistically summarize the correct and incorrect identification results for all samples, are displayed in Figs. 2(c) and 3(c) for Task I and Task II performed by the PM-DNN and WM-DNN, respectively. Numerical evaluations on the test datasets confirm that the dual-channel PM-DNN and WM-DNN successfully achieve parallel categorization of MNIST and FMNIST targets with high accuracy, reaching rates above 97% and 86%, respectively. While polarization multiplexing is limited to two orthogonal channels, wavelength multiplexing can support a large number of channels, thus allowing for higher-capacity computation. To assess the ability of metasurface-based M-DNNs to handle a greater number of tasks simultaneously, we construct a tri-channel WM-DNN to perform three-task classification: MNIST (Task I), FMNIST (Task II), and KMNIST, which contains 10 classes of kanji Japanese characters (Task III). The datasets for these tasks are encoded at wavelengths of 450 nm, 550 nm, and 650 nm, respectively, under x-polarized light. The bottom panel of Fig. 1(c) shows the detection plane used for the tri-channel WM-DNN, with specific sub-areas dedicated to specific tasks. Three individual single-task DNNs are trained at 450 nm, 550 nm, and 650 nm, tasked with recognizing MNIST, FMNIST, and KMNIST, respectively. The upper panel of Fig. 4(a) displays the phase maps corresponding to the hidden layers of the three single-task DNNs for Tasks I to III, obtained from training. The tri-channel metasurface-based WM-DNN is realized by locally assigning meta-atoms from our library, each meticulously selected to meet the three-dimensional phase requirement at each pixel. The errors are more substantial than the dual-channel cases, particularly for the FMNIST and KMNIST tasks. This is expected as increasing the number of tasks exacerbates task competition, making it challenging for the meta-atoms to meet all the phase modulation requirements. The classification accuracies yielded from assessing the tri-channel WM-DNN on the MNIST, FMNIST, and KMNIST test datasets are 96.73%, 80.9%, and 81.13%, respectively. As shown in Table 1, the identification accuracies for FMNIST and KMNIST are notably lower than those in their corresponding single-task DNNs, which aligns with the phase map results discussed earlier. Moreover, these two tasks are more challenging than MNIST and, therefore, more sensitive to errors.

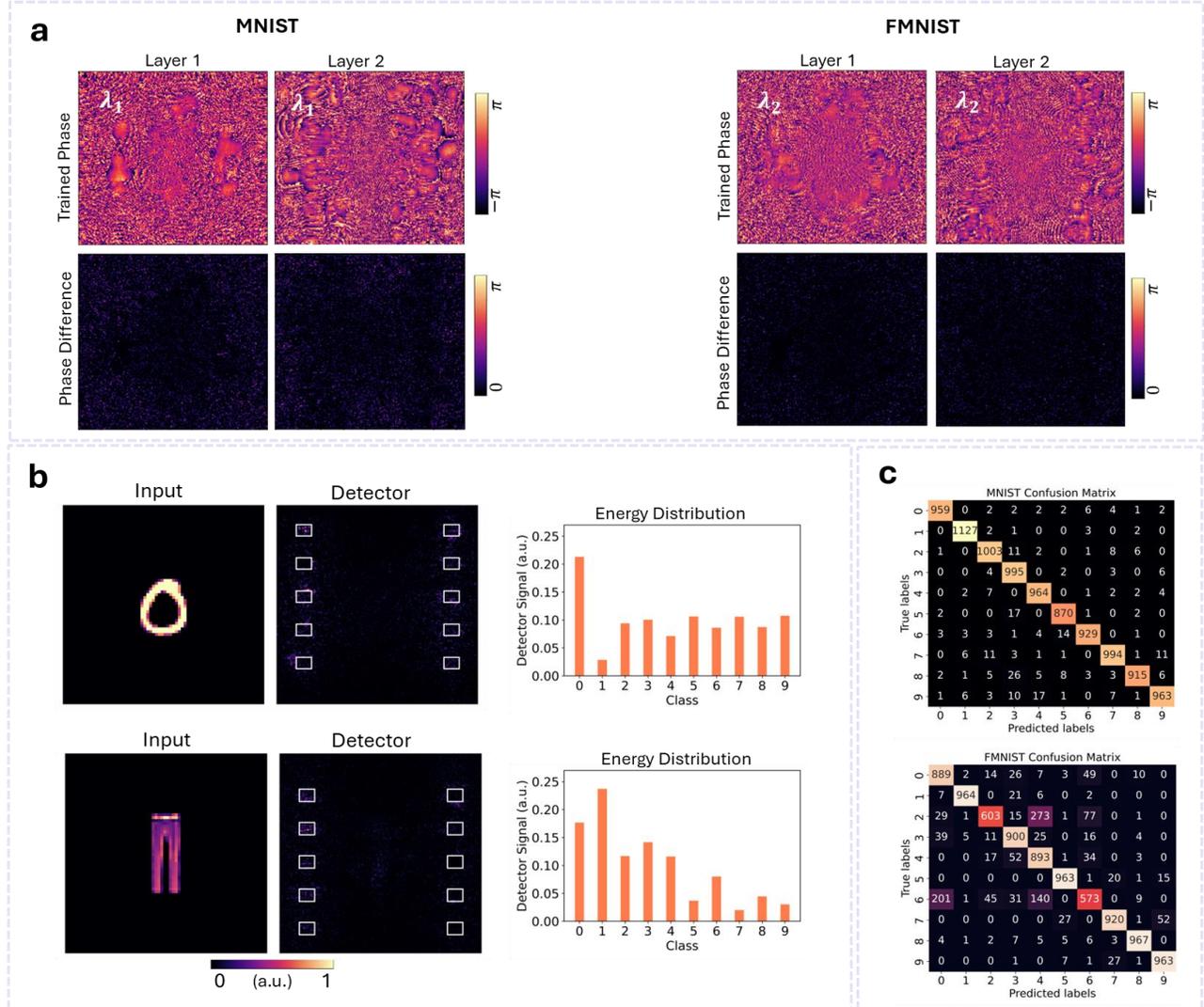

Fig. 3(a) (Top) Final phase profiles of the two hidden layers of the dual-channel WM-DNN under x-polarized incident light at 450 nm (left) and 550 nm (right). (Bottom) The absolute phase difference between the desired phase and the phase realized by the metasurfaces at each pixel. (b) Exemplary results of simultaneously classifying a handwritten digit and a fashion item, encoded at wavelengths of 450 nm and 550 nm, respectively. Output intensity patterns and normalized energy distributions across the category sub-regions illustrate the capability of the WM-DNN in parallel two-task classification. (c) Confusion matrices for MNIST and FMNIST processed by the WM-DNN, presenting the classifier's performance across individual classes, with average classification accuracies of 97.19% and 86.35% for Task I and Task II, respectively.

Fig. 4(b) depicts the performance of the tri-channel WM-DNN in parallel recognition, processing '2' from MNIST, ankle boot' from FMNIST, and 'お' from KMNIST. The detector intensity patterns and the normalized energy distributions of the sub-areas for different tasks indicate that the multi-wavelength incident light is successfully directed onto the correct sub-regions corresponding to each task. Fig. 4(c) presents the overall identification results across all samples from the three test datasets through confusion matrices, revealing that the average performance across all classes is commendable.

Table 1. Accuracy of the M-DNNs

| DNN Model | Design Method | Task I (MNIST) | | Task II (FMNIST) | | Task III (KMNIST) | |
|---|---|---|---|---|---|---|---|
| | | Single-task | Multi-task | Single-task | Multi-task | Single-task | Multi-task |
| Dual-Channel PM-DNN | Meta-atom library | 97.75% | 97.72% | 88.04% | 88.01% | - | - |
| Dual-Channel WM-DNN | Meta-atom library | 97.66% | 97.19% | 87.80% | 86.35% | - | - |
| Tri-Channel WM-DNN | Meta-atom library | 97.49% | 96.73% | 88.03% | 80.9% | 88.88% | 81.13% |
| Tri-Channel WM-DNN | End-to-end method | 97.49% | 96.48% | 88.03% | 85.68% | 88.88% | 85.35% |

It is worth discussing the extent of neuronal connectivity in our trained DNNs. Connectivity between layers is a crucial factor influencing the computational implementation complexity and the inference performance of the DNN. At a given wavelength, the interconnectivity between the neurons is dictated by the neuron size within each layer, which determines the diffraction angle, along with the axial distance between the layers. As it is analyzed in [22], for a DNN to be considered fully connected, the radius of the diffraction spot of each neuron $R$ should be larger than the side length of the diffractive layer. This radius is defined as: $R = z \times tan\varphi_{max}$, where $\varphi_{max} = sin^{-1}(\lambda/2d)$, with $\varphi_{max}$ representing the half-maximum-diffraction angle, $d$ the neuron size and $z$ the axial distance between successive layers.

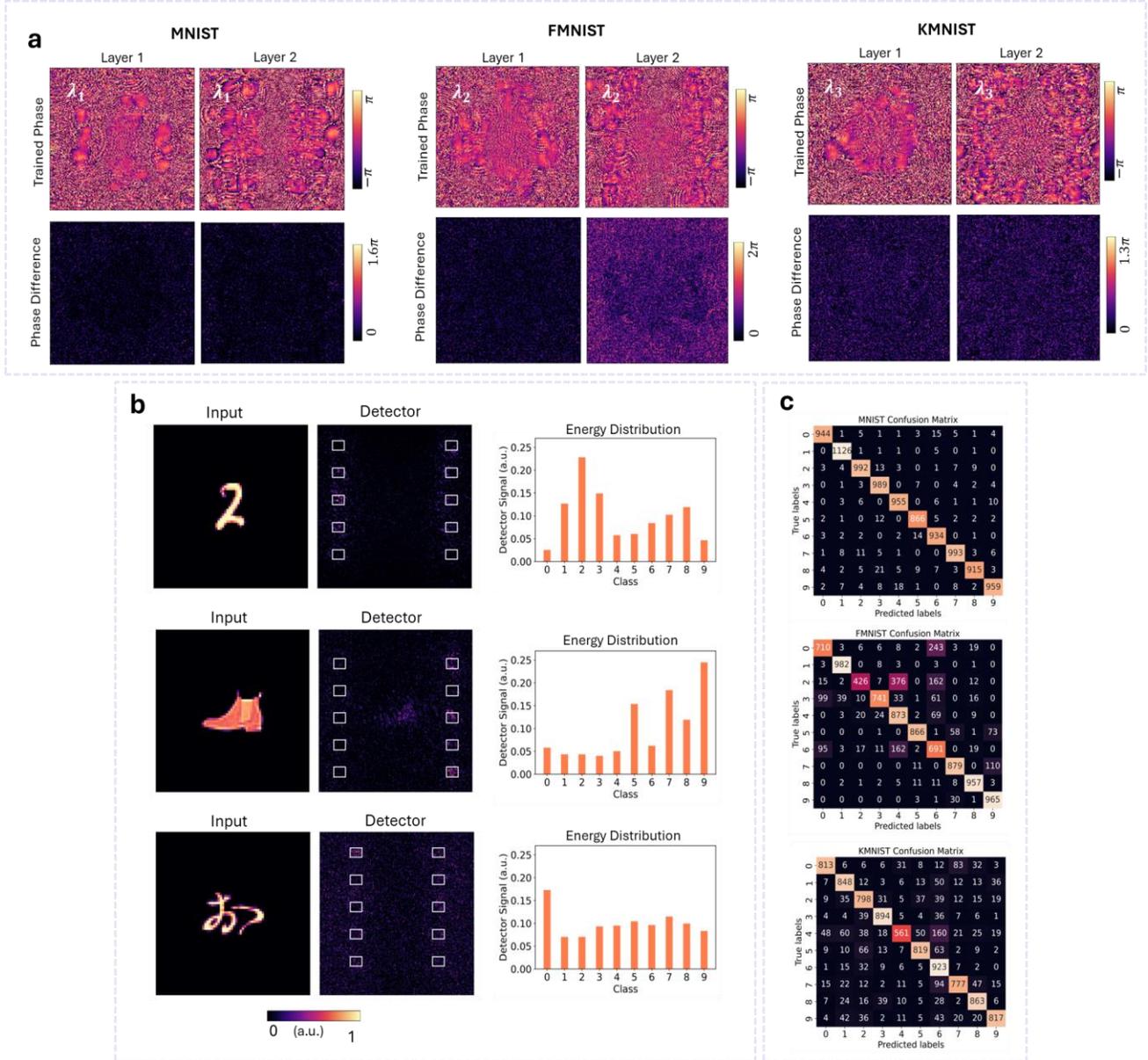

Fig. 4(a) (Top) Final phase profiles of the two hidden layers of the tri-channel WM-DNN with x-polarized incident light at 450 nm (left), 550 nm (middle), and 650 nm (right). (b) Exemplary results of simultaneously classifying a handwritten digit, a fashion item, and a Kanji character, encoded at wavelengths of 450 nm, 550 nm, and 650 nm, respectively. (Bottom) The absolute phase error between the optimized phase and the phase realized by the metasurfaces at each pixel. Output intensity patterns and normalized energy distributions across the category sub-areas demonstrate the effectiveness of the WM-DNN in performing parallel recognition of three tasks. (c) Confusion matrices for MNIST, FMNIST, and KMNIST processed by the WM-DNN, summarizing the classifier's performance across individual classes, with average classification accuracies of 96.73%, 80.9%, and 81.13%, for TaskI, Task II, and Task III, respectively.

Accordingly, the diffraction radii for the neurons in our designed DNNs at wavelengths of 450 nm, 550 nm, and 650 nm are 140 μm, 175 μm, and 212 μm, respectively. These values are smaller than the side length of the layers, which is 252 μm, thus making our DNNs partially connected. Nevertheless, although our DNNs are not fully connected neural networks, the large coverage of the diffraction spots still provides sufficient connectivity for effective information processing. This is evidenced by the training results of our single-task DNNs (as presented in Table 1), which achieved accuracies nearly comparable to those of fully connected networks reported in the literature [10, 45]. Moreover, it is important to note that, in the design of a multiwavelength DNN, despite the fixed layer spacing and neuron size across different channels, channels with higher wavelengths exhibit greater connectivity due to larger diffraction angles, enhancing computational capability. As such, assigning more demanding tasks to channels with higher wavelengths is beneficial; for instance, we assigned MNIST, the least complex task, to our channel with the lowest wavelength.

## 4. Designing M-DNNs using an end-to-end approach

Lastly, we investigate an end-to-end design strategy to address issues encountered in the previous examples, with the imperfect phase map implementation using the meta-atom library approach. Unlike the previous approach, where the DNN for each task was optimized separately, we now incorporate a joint optimization framework to minimize classification errors across all three tasks simultaneously. By constructing surrogate models that relate the structural parameters of the meta-atoms to their channel-dependent complex transmission responses and integrating these models as proxy functions during training, we can directly optimize the structural parameters of the meta-atoms. Therefore, in contrast to the approach in the previous section, the trainable parameters of the diffractive networks here are the structural parameters of the meta-atoms, rather than their transmission responses. This approach eliminates the need for a further search step in the design process, thus addressing the issues caused by the imperfect phase map realization. In addition, it ensures that the training process accounts for the physical constraints of the unit cells, producing feasible and physically realizable complex modulation responses at each channel, thereby enhancing the performance of M-DNNs implemented by the metasurfaces.

As a final demonstration, we implement a tri-channel WM-DNN to simultaneously perform MNIST, FMNIST, and KMNIST tasks using similar wavelength channels and sub-area designations on the detection plane, as in the previous design. For the surrogate models that map the widths of the meta-atoms to their optical modulation coefficients at each wavelength, we utilize deep Artificial Neural Networks (ANNs) as differentiable proxy functions. The use of deep learning in nanophotonics is a well-established practice and has proven highly effective, particularly in inverse design, as evidenced by references [47-53].

The architecture used for the ANN models, depicted in Fig. 5(a), consists of four hidden layers, each with 512 neurons and a ReLU activation function, along with two input neurons representing $w_x$ and $w_y$, and three output neurons representing transmittance, and the cosine and sine of the phase shift. We design three models, $f_1(w_x, w_y)$, $f_2(w_x, w_y)$, and $f_3(w_x, w_y)$, using this architecture to approximate the transmission responses of the nanofins under the x-polarized light at wavelengths of 450 nm, 550 nm, and 650 nm, respectively. For each wavelength, 2784 cell

instances—sourced from our previous COMSOL simulations—are used for training, while a separate set of 696 unseen cell instances is reserved for testing. The assessment of the unseen dataset shows that, while there are variations in performance across different models, they all exhibit low prediction errors, as presented in Table 2. As a representative example, Fig. 5(b) shows the progress of training and test losses for the model $f_1(w_x, w_y)$ trained at 450 nm over 2500 epochs, reaching a mean squared error (MSE) of 0.047 on the test samples. Fig. 5(c-e) provide a visual comparison between the transmission responses of the meta-atoms derived from the COMSOL simulations and those predicted by the ANN models for all three wavelengths, verifying that the ANN models can serve as reliable approximation functions.

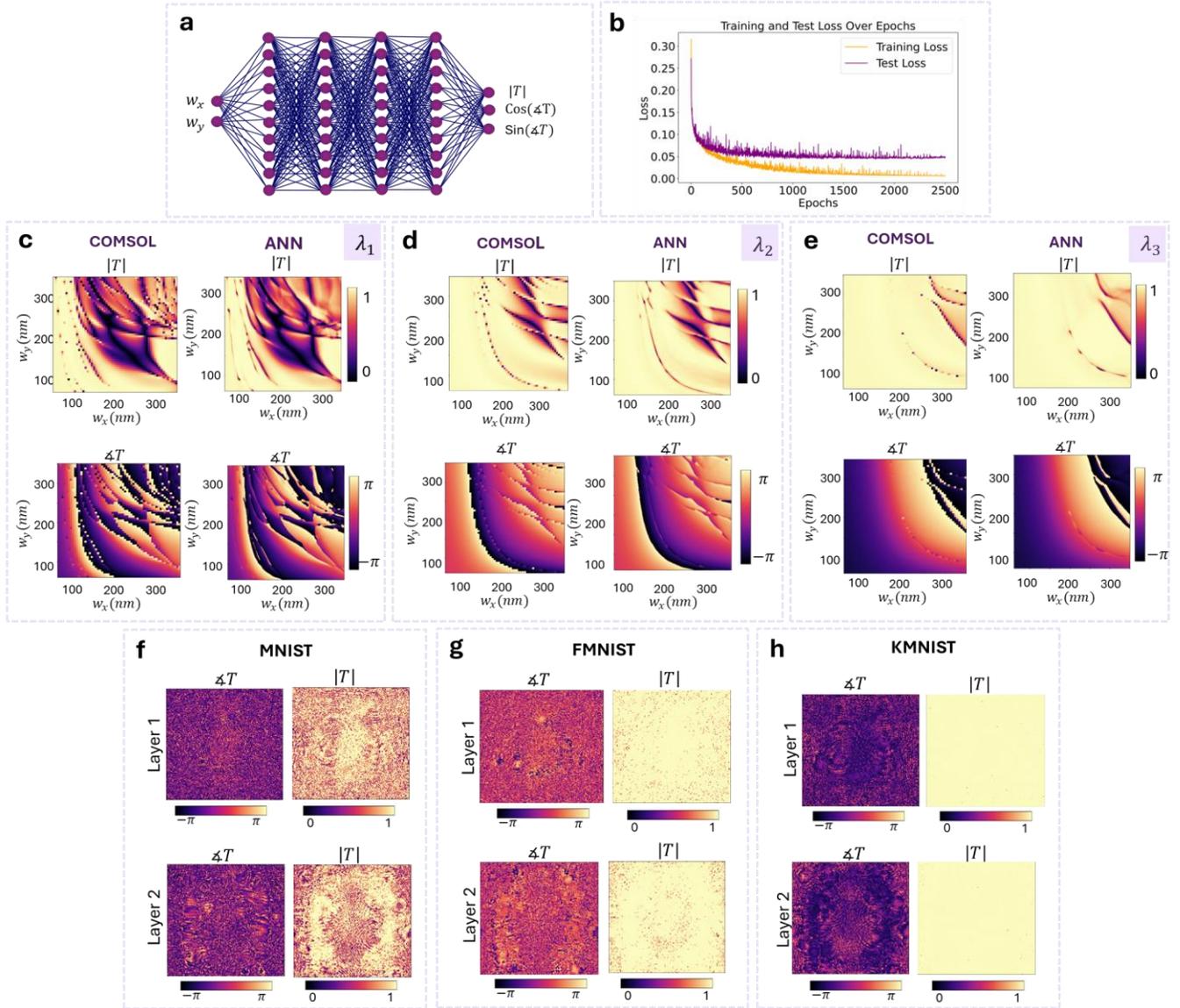

Fig. 5(a) The architecture of the ANN models, which are used to map the phase and transmittance values of the meta-atoms to their structural parameters. (b) Training and test losses for the ANN model trained at 450 nm, over 2500 epochs, illustrating the network's convergence to final training and test losses of 0.0056 and 0.0471, respectively. (c-e) Comparison of COMSOL and ANN outputs for different under x-polarized light at 450 nm, 550 nm, and 650 nm. Data points from COMSOL simulations used for training are evaluated at 5 nm intervals, while the ANN outputs are at a resolution of 1 nm. (f-h) Final amplitude and phase modulation profiles for the two hidden layers of the tri-channel WM-DNN obtained through joint training, utilizing the pre-trained models from parts (c-e).

Once trained, the ANN models have their network weights fixed and are then employed to design our tri-channel WM-DNN. In each iteration of training, the structural parameters of the meta-atoms are updated based on the back-propagation algorithm, with the models predicting their associated complex modulation coefficients, thereby facilitating the simultaneous training of multiple tasks. The complex transmittance maps of the hidden layers obtained from the joint training session are displayed in Fig. 5(f, g, and h), respectively.

Table 2. Training and Test Losses for ANN Models Used in the Inverse Design Process

| ANN Model | Training Loss (MSE) | Test Loss (MSE) | Number of Epochs |
|---|---|---|---|
| $f_1(w_x, w_y)$ | 0.0056 | 0.0471 | 2500 |
| $f_2(w_x, w_y)$ | 0.0019 | 0.0207 | 1500 |
| $f_3(w_x, w_y)$ | 0.0037 | 0.0074 | 1500 |

It's worth noting that compared to previous cases where the networks were trained with a single objective, the learning process in this scenario faces a more complex optimization landscape, primarily due to the multi-objective joint training. Additionally, the modulation coefficient of the unit cell has a more intricate relationship with the trainable parameters $w_x$ and $w_y$, represented as $|f_i| exp\left(i \sphericalangle f_i(w_x, w_y)\right)$. This complexity is particularly pronounced at 450 nm, where resonances and fluctuations are most significant. In contrast, the previous phase-only scenario involved a simpler relationship, where the learnable parameter $\varphi$ (i.e., the transmission phase) had a straightforward exponential relationship with the modulation coefficient, $exp(i\varphi)$, and the amplitude was set to unity. As a result, more time and computational resources are required for the joint network to converge. To facilitate better convergence of the joint training, we utilize a weighted sum loss function:

$$Total\ loss = w_1 * loss_{MNIST} + w_2 * loss_{FMNIST} + w_3 * loss_{KMNIST} \qquad (2)$$

We conduct multiple training sessions with various weight combinations, and the best results are achieved with coefficients $w_1$, $w_2$ and $w_3$ set at 0.13, 0.2, and 0.67, respectively. Utilizing a genetic algorithm to optimize these weights can further enhance the model's performance, though it would involve higher computational costs.

Blind testing of the tri-channel WM-DNN trained with our end-to-end method demonstrates accuracies of 96.48% for MNIST, 85.68% for FMNIST, and 85.35% for KMNIST. Compared to the previous tri-channel WM-DNN designed using the meta-atom library approach, these results reproduce the high accuracy for MNIST and show improvements of 4.7% and 4.2% for FMNIST and KMNIST, respectively, highlighting the effectiveness of this method. Fig. 6(a) illustrates the performance of the end-to-end designed tri-channel WM-DNN in simultaneously classifying '1', 'trouser', and 'お' from the MNIST, FMNIST, and KMNIST datasets, respectively. The system successfully predicts the correct category for each task. The confusion matrices, shown in Fig. 6(b), display the network's performance across all individual classes for all test samples in the three tasks. A comparison with the confusion matrices in Fig. 4(c) reveals an increase in the total number of correct identifications, confirming the improvement in average accuracies. For instance, in the KMNIST dataset, the number of correct identifications for class 4—the worst-performing class—increased from 561 to 842, demonstrating a clear improvement in the recognition of KMNIST images.

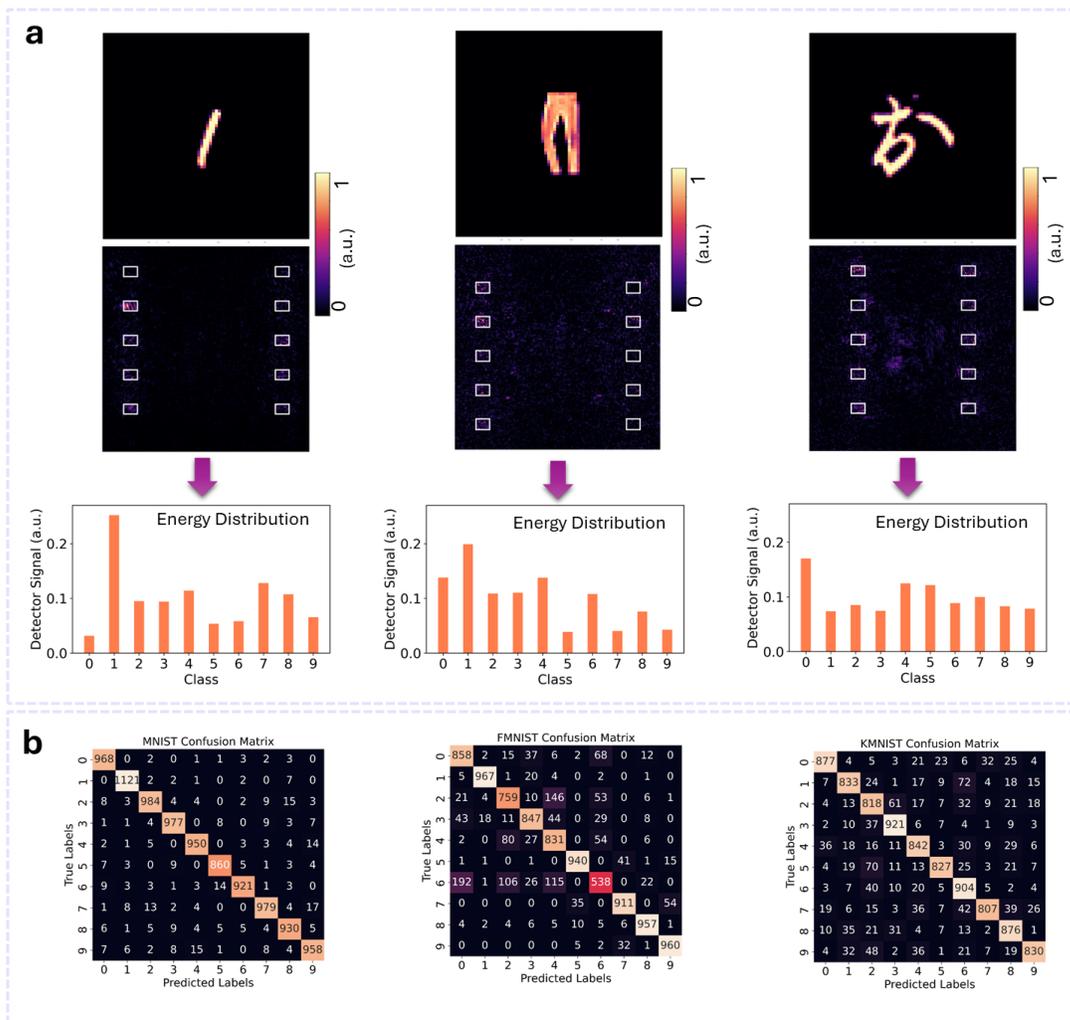

Fig. 6. (a) Exemplary results of simultaneously classifying a handwritten digit (left), a fashion item (middle), and a Kanji character (right), encoded at wavelengths of 450 nm, 550 nm, and 650 nm, respectively, and processed by the jointly trained tri-channel WM-DNN. Output intensity patterns and normalized energy distributions across the category sub-areas confirm the capability of the WM-DNN for simultaneous three-task classification. (b) Confusion matrices for MNIST, FMNIST, and KMNIST processed by the WM-DNN, summarizing the classifier's performance across individual classes, with average classification accuracies of 96.48%, 85.68%, and 85.35%, for TaskI, Task II, and Task III, respectively, showing improvement compared to the previous WM-DNN design with the meta-atom library approach.

# 5. Conclusion

In this work, we have demonstrated the potential of wavelength and polarization multiplexing schemes to facilitate all-optical multi-task learning using DNNs with bilayer cascaded metasurfaces. We design dual-channel PM-DNN and WM-DNN, utilizing polarization and wavelength multiplexing, respectively, using our meta-atom library. Numerical results for both systems in parallel processing of two tasks show high accuracies comparable to their individually trained single-task counterparts, thereby validating the effectiveness of our dual-channel M-DNN design methodology. Additionally, we explore the implementation of a tri-channel WM-DNN to perform three classification tasks in parallel, using two different design strategies: the meta-atom library approach and the end-to-end joint training framework. While the meta-atom library approach achieves high accuracy for the MNIST task, it exhibits a moderate decline in performance for the FMNIST and KMNIST tasks. This decline can be attributed to the increased number of tasks, which makes finding meta-atoms that satisfy all the local modulation coefficient requirements more challenging.

On the other hand, the end-to-end joint training framework uses approximate models to map the structural parameters of the unit cells to their complex transmission responses for each channel, ensuring that the obtained modulation coefficients are realizable by the meta-atoms, thereby enhancing the multi-tasking performance of the network. However, despite the improvements, the performance of the M-DNN still falls short when compared to the individually trained DNNs for the respective single tasks. We anticipate that performance could be further improved by incorporating more complex meta-atom geometries, increasing the number of layers, and applying hyperparameter tuning, possibly using a genetic algorithm, to the joint training. Finally, it is worth noting that the significance of the end-to-end training approach becomes increasingly apparent as the number of parallel tasks grows, offering considerable advantages for designing highly parallel optical machine learning systems capable of handling numerous tasks simultaneously. Such systems could be further utilized in high-throughput computational imaging, real-time data processing, and autonomous systems.

## Materials and Methods

**Training of the Neural Networks**

All neural networks, including DNNs and ANNs, were implemented using Python version 3.10 and TensorFlow framework version 2.16.1 on Google Colab Pro, equipped with a V100 NVIDIA GPU and 32 GB of RAM. We used the MNIST, FMNIST, and KMNIST datasets, with each image—originally a 28×28 grayscale image—zero-padded to 70×70 and then resized to 210×210. These images were encoded in the amplitude of light for the corresponding channels. Each dataset contains 60,000 training samples and 10,000 testing samples across 10 classes. The Adam optimizer was used to train all networks, with an exponential decay learning rate schedule starting at 0.001 and a decay rate of 0.9 applied every 10,000 steps, using a batch size of 32. A cross-entropy loss function was employed to maximize light intensity in the target region for the DNN designs.

**Meta-atom Search and Selection Process**

As described in the paper, the proposed M-DNNs outlined in Section 3 were implemented through a comprehensive search within our unit cell library. Specifically, to select the optimal meta-atom at each location, we first excluded meta-atoms with transmission amplitudes below 0.5. We then applied a weighted sum of the errors between the target phases and the phases obtained from the meta-atoms across all tasks as our search criterion, selecting the meta-atoms that minimized this total error. Utilizing the weighted sum error as the search metric enables us to prioritize the more challenging and sensitive tasks, thereby enhancing the overall multitasking performance. We conducted several iterations of searches with varying weight combinations to find the optimal set of weights that achieves accuracies greater than 80% for all tasks, rather than maximizing the accuracy for one task at the expense of weaker performance in the others. The optimal error weights for the metasurface implementation of the dual-channel PM-DNN were identified as 1 and 1.2 for Task I and Task II, respectively, while for the dual-channel WM-DNN, weights of 1.2 and 2 for Task I and Task II yielded the desired accuracies across both tasks. Lastly, for the tri-channel WM-DNN, the optimal weights were determined to be 0.8, 1.28, and 1.23 for Task I, Task II, and Task III, respectively.

## Acknowledgment

This work is supported by the National Science Foundation Award number ECCS-2240448.